\documentstyle[twocolumn,psfig]{article}
\begin{document}

\renewcommand{\arraystretch}{1.0}
\renewcommand{\textfraction}{0.05}
\renewcommand{\topfraction}{0.9}
\renewcommand{\bottomfraction}{0.9}
\def\la{\;\raise0.3ex\hbox{$<$\kern-0.75em\raise-1.1ex\hbox{$\sim$}}\;}
\def\ga{\;\raise0.3ex\hbox{$>$\kern-0.75em\raise-1.1ex\hbox{$\sim$}}\;}
\def\lr{\;\raise0.3ex\hbox{$\rightarrow$\kern-1.0em\raise-1.1ex\hbox{$\leftarrow$}}\;}
\newcommand{\kB}{\mbox{$k_{\rm B}$}}           
\newcommand{\tn}{\mbox{$T_{{\rm c}n}$}}        
\newcommand{\tp}{\mbox{$T_{{\rm c}p}$}}        
\newcommand{\te}{\mbox{$T_{eff}$}}             
\newcommand{\dash}{\mbox{--}}                  
\newcommand{\ex}{\mbox{\rm e}}                 
\newcommand{\r}{\rule{0cm}{0.4cm}}
\newcommand{\hh}{\rule{0.5cm}{0cm}}
\newcommand{\hb}{\rule{0.4cm}{0cm}}
\newcommand{\rl}{\rule{0em}{1.8ex}}
\newcommand{\hhh}{\rule{1.4em}{0ex}}
\newcommand{\hhb}{\rule{1.2em}{0ex}}
\newcommand{\hhl}{\rule{2.4em}{0ex}}
\newcommand{\s}{$\;\;$}

\title{\bf Electric Character of Strange Stars
        }
\author{R.X.~Xu, G.J.~Qiao \\
        National Astronomical Observatories, BAC and Astronomy Department, \\
	Peking University, Beijing 100781, China
	\footnote{BAC is CAS-PKU joint Beijing Astrophysical Center}
        \footnote{e-mail: RXXU@bac.pku.edu.cn}\\
  \\
\Large{\it Published in Chin.Phys.Lett., Vol.16, p.778}
}
\date{}
\maketitle

\begin{abstract}

Using the Thomas-Fermi model, we investigated the electric 
characteristics of a static non-magnetized strange star without crust 
in this paper. The exact solutions of electron number density and 
electric field above the quark surface are obtained. These results 
are useful if we are concerned about physical processes near the quark 
matter surfaces of strange stars.

\end{abstract}

\bigskip
\bigskip

If strange quark matter in bulk is absolutely stable, there might be 
strange star [1] consisting almost completely of strange quark matter 
in the universe. Frustratingly, strange stars are very similar to 
neutron stars in their many properties, such as mass and radius. Thus, 
it is suggested that pulsars might be strange stars [1-3]. However, 
the interesting question about the nature of pulsars (neutron stars 
or strange stars) has not been answered with certainty even now yet.

Strange quark matter mainly consists of up, down, and strange quarks. 
As strange quark is a little more massive than that of up and down 
quarks, there are a few electrons in the chemical equilibrium of 
strange quark matter in order to keep the matter neutral. Hence, 
electromagnetic interaction as well as strong interaction results in 
strange quark matter. The electromagnetic force participated in makes 
the structure of strange quark matter more interesting and attractive. 
In this paper, we are to investigate this electric peculiarity of 
strange stars. Previously, some numerical results [1,4] have been 
given in literature, but no exact analytical result appears.

For a static and non-magnetized strange star, the properties of 
strange quark matter are determined by the thermodynamic potentials 
$\Omega_i$ (i = u, d, s, e) which are functions of chemical potential 
$\mu_i$ as well as the strange quark mass, $m_s$, and the strong 
interaction coupling constant $\alpha_c$ [1,5]. We use units where 
$\hbar=c=1$, physical quantities can be changed to be expressed in 
units of c.g.s. by using $\hbar c = 197.327$ fm$\cdot$MeV and $c=2.9979\times 
10^{10}$ cm/s. Assuming weak interaction chemical equilibrium and 
overall charge neutrality, we come to
$$
\begin{array}{lll}
\mu_d & = & \mu_s = \mu,\\
\mu_e & + & \mu_u = \mu,\\
n_e & = & (2n_u - n_d - n_s)/3,\\
n_i & = & -{\partial \Omega_i \over \partial \mu_i},
\end{array}     \eqno(1a)
$$
and the total energy density $\rho$ reads
$$
\rho = \sum_{i=1}^4 (\Omega_i+\mu_i n_i)+B,     \eqno(1b)
$$
where $B$ is the bag constant, and $\Omega_i$ referred to the Appendix 
in the paper by Alcock et al.[1]. The above equations (1a-1b) have only 
one free independent parameter, $\mu$, and establish the relations 
for $\rho, \mu_i, n_i$ (i = 1,2,3,4 for u, d, s, e, respectively).
There are nine equations for these nine variables; therefore equ. 
(1a-1b) are self-contained.

The calculation results from equ.(1a-1b) are shown in Fig. 1,
   \begin{figure*}
      \psfig{figure=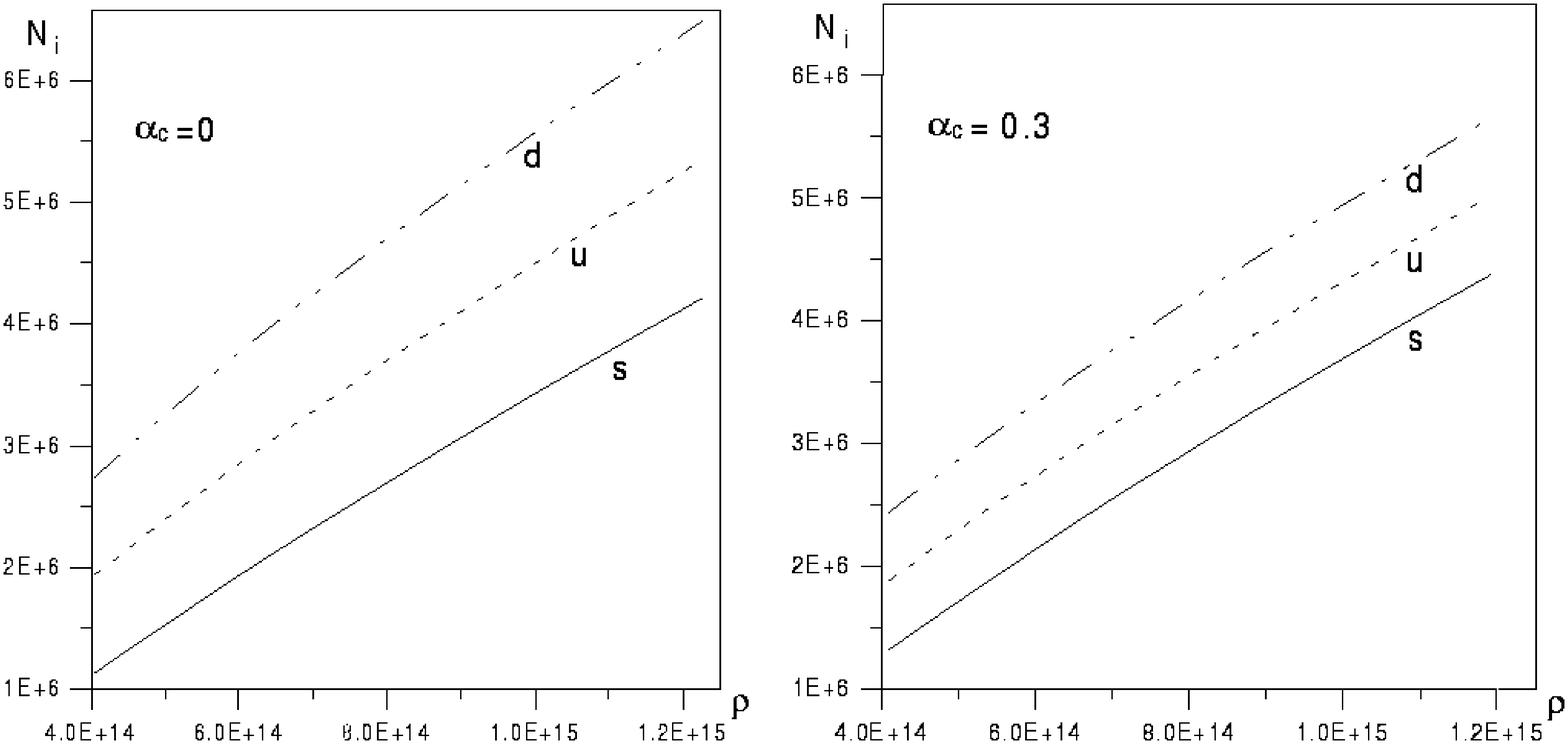,width=17.5cm,angle=-0}
      \parbox[b]{17.5cm}{\caption[]{
    The number densities of u, d, and s quarks, $N_u$,
       $N_d$, $N_s$, are functions of total energy density $\rho$
       (in g/cm$^3$). $N_i$ refers to one of $N_u$, $N_d$, and $N_s$,
       which are in unit of particle number per cm$^3$. The couple
       constant $\alpha_c$ is chosen to be 0 (left) and 0.3
       (right), respectively.
              }}%
         \label{FigGam}%
    \end{figure*}
and 2, where the number densities of u, d, s quarks, and the quark 
charge density $\rho_q$ (in unit of Coulomb per cm$^3$) are varied 
as a function of total energy density $\rho$. In the computation, we 
choose $B = (145 MeV)^4$, $m_s = 200$ MeV, and the renormalization 
point $\rho_R = 313$ MeV, both for $\alpha_c = 0$ and $\alpha_c = 0.3$. 
As $\rho$ has a mild rise variation from the outer part to the interior 
of a strange star [1], the number density of u, d, and s quarks increase 
almost in a same degree. However, the equilibrium quark charge density 
$\rho_q$ changes significantly, as $\rho$ increases (Fig. 2),
%
   \begin{figure}
      \psfig{figure=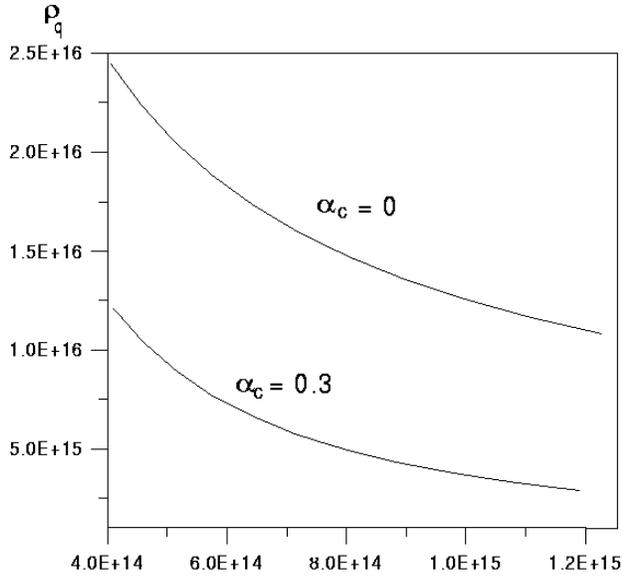,width=8.8cm,angle=0}
      \caption[]{
The quark charge density $\rho_q$ (in unit of Coulomb per cm$^3$) 
decreases 
as a function of the total energy density $\rho$. The coupling 
constant 
$\alpha_c$ is chosen to be 0 and 0.3, respectively.
              }
         \label{FigVibStab}
   \end{figure}
which means the number of equilibrium electrons becomes smaller as 
one goes to a deeper region of a strange star. For a strange star with 
a typical pulsar mass $1.4 {\rm M}_\odot$, the total energy $\rho$ has a 
very modest variation with radial distance of strange star [1], from 
$\sim4\times10^{14}$g cm$^{-3}$ (near surface) to $\sim 7\times 
10^{14}$g cm$^{-3}$ (near center), therefore the quark charge density 
$\rho_q$ would be order of $10^{15}$ ($\alpha_c = 0.3$) to $10^{16}$ 
($\alpha_c = 0$) Coulomb cm$^{-3}$. Physically, as the Fermi energy 
of quarks becomes higher (for larger $\rho$), the effect due to $m_s 
\not= 0$ would be less important, hence, the charge density should 
be smaller. For a rotating magnetized strange star with typical 
parameters of pulsars, the Goldreich-Jullian space charge separated 
density is very small (in order of $10^{-7}$ Coulomb/cm$^3$[6]),
so it is a very 
good approximation to neglect the space charge separation, i.e., the 
calculation results in Fig.1 and 2 are valid for rotating 
magnetized strange stars.

Since the quark matter are bounded through strong interaction (the 
thickness of the quark surface will be of order 1 fm), and the 
electrons are held by the quark matter electrically, hence the 
electron's distribution would extend beyond the quark matter surface. 
A simple Thomas-Fermi model has been employed to solve for this 
distribution [1], and the local charge distribution can be obtained 
by Poisson's equation
$$
{d^2 V\over dz^2} =
\left\{    \begin{array}{ll}
{4\alpha\over 3\pi}(V^3-V_q^3) & z\leq 0,\\
{4\alpha\over 3\pi} V^3 & z > 0,
\end{array}     \right.  \eqno(2)
$$
where z is a measured height above the quark surface, $\alpha$ is the 
fine-structure constant, $V_q^3/(3\pi^2)$ is the quark charge 
density, $V/e$ is the electrostatic potential, and the number density 
of electrons is given by
$$
n_e = {V^3\over 3\pi^2}. \eqno(3)
$$
Physically, the boundary conditions for equ.(2) are
$$
\begin{array}{ll}
z \rightarrow -\infty: & V \rightarrow V_q, dV/dz
\rightarrow 0;\\
z \rightarrow +\infty: & V \rightarrow 0,   dV/dz
\rightarrow 0.
\end{array}
$$
By a straightforward integration of equ.(2), without considering the 
boundary conditions, we can get
$$
{dV \over dz} = 
\left\{    \begin{array}{l}
-\sqrt{{8\alpha\over 3\pi} (V^4/4 - V_q^3 V) + C_1},  (z<0)\\
-\sqrt{{2\alpha\over 3\pi} V^4 + C_2}.  (z>0)
\end{array}     \right.
$$
where, $C_1$ and $C_2$ are two constants determined by the boundary 
conditions. Using the first condition, we get $C_1 = {2\alpha \over 
\pi}V_q^4$. Using the second one, we get $C_2=0$. Therefore, we come 
to
$$
{dV \over dz} = 
\left\{    \begin{array}{l}
-\sqrt{2\alpha\over 3\pi} \cdot \sqrt{V^4 - 4 V_q^3 V + 3 V_q^4},  
(z<0)\\
-\sqrt{2\alpha\over 3\pi} \cdot V^2. (z>0)
\end{array}     \right.  \eqno(4)
$$
The continuity of equ. (4) at $z=0$ educes the result $V(z=0) = 3V_q/4$ 
[1], and we can consider the solution of equ.(4) for $z > 0$ by
$$
{dV \over dz} = -\sqrt{2\alpha\over 3\pi} V^2,  {\rm
boundary}: V(z=0) = {3\over 4}V_q,      \eqno(5)
$$
hence,
$$
V={3V_q\over \sqrt{6\alpha\over\pi}V_qz+4},\;\;({\rm
for} z > 0). \eqno(6)
$$
Therefore the number density of electrons is [from equ. (3)]
$$
\begin{array}{lll}
n_{\rm e} & = & {9V_q^3\over \pi^2 
(\sqrt{6\alpha\over\pi}V_qz+4)^3}\\
& \sim & {9.49 \times 10^{35} \over (1.2 z_{11} + 4)^3}
\;\; {\rm cm^{-3},}
\end{array} \eqno(7)
$$
and the electric field reads
$$
\begin{array}{lll}
E = -{dV \over dz} & = & \sqrt{2\alpha\over 3\pi} \cdot
{9 V_q^2 \over
    (\sqrt{6\alpha\over \pi} V_q \cdot z + 4)^2}\\
& \sim & {7.18 \times 10^{18} \over (1.2 z_{11} + 4)^2}
\;\; {\rm V\;cm^{-1},}
\end{array}     \eqno(8)
$$
where, the direction of the electric field is outward, and $V_q$ has 
been chosen to be 20 MeV (hence $\rho_q = 270.19$ MeV$^3 = 
5.63\times10^{15}$ Coulomb cm$^{-3}$), $z_{11} = z/(10^{-11}$ cm).

It is interesting that, from equ.(8), although the electric field near
the surface is about $10^{17}$ V cm$^{-1}$, the electric field 
decreases very quickly above the quark surface. The calculation of 
electric field shows that the electric field is $\sim 10^{11}$ V 
cm$^{-1}$ when $z\sim 10^{-8}$ cm, which means the induced electric 
field should be dominant when $z > 10^{-8}$ cm and bare strange stars 
(i.e. without crust) can have magnetospheres [3]. Also the electron 
charge density calculated from equ.(7) decreases from $2.4\times 
10^{15}$ Coulomb cm$^{-3}$ (where $V = {3\over 4}V_q $) at the surface 
to $3.3\times 10^{-9}$ Coulomb cm$^{-3}$ when $z = 3\times 10^{-3}$ 
cm, while the Goldreich-Jullian space charge separated density is in 
order of $10^{-8}$ near strange stars surface.

Equ.(7) and (8) are effective in discussing the properties of strange 
stars near the quark surface. Two examples are given here.

{\bf Example 1:} As a strong outward-directed electric field exists 
near the quark surface, accreted matter will be repulsed just above 
the surface because of Coulomb force; hence a crust around the strange 
quark core might be formed. The properties of this crust have 
sufficiently been discussed in literature [1,4,7], but no simple 
relation to describe the crust mass and the electric gap width. Since 
equ. (7) can be used to study the behavior of a test charged point, 
by introducing a coefficient $\eta \sim 1$ denoting the effective 
(positive) electric charge, we can get a simple relationship of crust 
mass $M_{\rm crust}$ and electric gap width $z_{\rm G}$ for typical 
crust and strange star values
$$
M_{\rm crust} \sim {29 \times 10^{-5} \over (1.2 z_{\rm G} + 4)^2} 
\eta \;\;\;{\rm M_\odot},
$$
where $z_{\rm G}$ is in $10^{-11}$ cm. For $z_{\rm G} \ll 1$ (the 
length scale of strong interaction is $\sim z_{\rm G}=10^{-2}$), we 
get the maximum values of crust mass $M_{\rm max} \sim 1.8 \times 
10^{-5} \eta {\rm M_\odot}$, which consists with previous results.

{\bf Example 2:} It is said that the Ruderman-Sutherland's inner-
gap model [8] has a `user friendly' nature for explaining the observed 
emission properties of radio pulsars. However, the RS model has a 
strange virtue: it can not be applied to half of neutron stars. It 
is assumed in RS model that the magnetospheric charge density above 
the polar cap is positive, which means the rotational angle velocity 
${\bf \Omega}$ and the magnetic momentum ${\bf \mu}$ are anti-
parallel (i.e. ${\bf \Omega}\cdot{\bf B}<0$). For neutron stars at 
which  ${\bf \Omega}$ and  ${\bf \mu}$ are parallel, which are called 
`anti-pulsars' (i.e. ${\bf \Omega}\cdot{\bf B}>0$), the inner gap can 
not be formed and the inner gap model does not work. If radio pulsars 
are bare strange stars [3], this strange virtue does not exist again. 
It is argued below that inner-gap sparking can also exist for 
anti-pulsars if radio pulsars are strange stars without crusts.

When ${\bf \Omega}\cdot{\bf B}>0$, the main reason that inner-gap 
does not exit is that neutron star can supply infinity of charged 
electrons, because electrons in the neutron stars' 
crusts can move across the magnetic field lines. (Electrons
can move across magnetic fields if their kinematic energy 
density $\gg$  the magnetic energy density.)
But for bare strange stars, the charge electrons 
available to be pull out from polar-cap are limited, and the times 
scale to pull out all these electrons is about $10^{-5}$ s. 
Therefore, polar-gap sparking can also be there for bare strange stars. 
The polar-cap area $S_{\rm c}$ reads
$$
\begin{array}{lll}
S_{\rm c} & = & {2\pi^2 R^3\over c P}\\
& \sim & 6.58 \times 10^8 R_6^3\cdot P_1^{-1}
\;\; {\rm cm^2},
\end{array}
$$
where $R$ is the radius of pulsars, $R_6=R/(10^6{\rm cm})$, $P$ is 
period, $P_1=P/(1{\rm s})$, and $c$ is the light speed. From equ. (7), 
the number of total electrons available to be pull out for a bare 
strange star is $Q_{\rm ss}$,
$$
\begin{array}{lll}
Q_{\rm ss} & = & S_{\rm c}\cdot e \int_{Z_{\rm c}}^{+\infty} n_{\rm 
e} {\rm d}z\\
& \sim & {7.25\times 10^{14}\over (1.2 z_{\rm c}+4)^2}
\;\; {\rm Coulomb},
\end{array}
$$
where $z_{\rm c}$ is a critical height in $10^{-11}$ cm, and $e$ is 
the elementary charge. At the height of $z_{\rm c}$, the electric 
field from quark matter and the induced unipolar electric field are 
equal (i.e, electrons are not forceful there). As $z_{\rm c}\sim 6\times 
10^3$ [3], $Q_{\rm ss}\sim 4.2\times 10^7$ Coulomb. Therefore, the 
time scale to pull out all these electrons is $\tau_{\rm ss}$
$$
\begin{array}{lll}
\tau_{\rm ss} & = & {Q_{\rm ss}\over \rho_{\rm GJ} S_{\rm c} c}\\
& \sim & 2.1\times 10^{-5}
\;\; {\rm s},
\end{array}
$$
where $\rho_{\rm GJ}\sim 10^{-7}$ Coulomb cm$^{-3}$ is the 
Goldreich-Jullian [6] space charge 
separated density at polar-cap. Recently, by studying the clear 
features of drifting pulses from PSR B0943+10, Rankin et. al. [9] find 
that the regular patterns of pulses consist with RS model, but the 
pulsar might be an anti-pulsar which is contrary to the expectations 
of RS model. We argue that PSR B 0943+10 is a bare strange star with 
${\bf \Omega}\cdot{\bf B}>0$.

\bigskip
{\it 
We would like to thank our pulsar group for helpful discussions.}

\end{document}